# The first attempts to measure light deflection by the Sun

Luís C. B. Crispino[1*] and Santiago Paolantonio[2]

*Soon after Einstein's calculation of the effect of the Sun's gravitational field on the propagation of light in 1911, astronomers around the world tried to measure and verify the value. If the first attempts in Brazil in 1912 or Imperial Russia in 1914 had been successful, they would have proven Einstein wrong.*

A hundred years ago, during the total solar eclipse of May 29, 1919, the photographs that confirmed for the first time the light deflection by the Sun's gravitational field were obtained. After a careful analysis of those photographic plates, the triumph of Einstein's theory of General Relativity over Newtonian gravitation was announced in a joint meeting of the Royal Society and the Royal Astronomical Society in November 6, 1919 [1, 2]. However, the 1919 eclipse was not the first occasion in which an attempt to measure light deflection by the Sun has been performed.

Back in 1911, Albert Einstein concluded that light rays would be bent by the Sun's gravitational field and that this could, in principle, be measured during a total solar eclipse [3]. The possibility of light bending has been conjectured long before Einstein, though. In the first query of his Optics Treatise, published in 1704, Isaac Newton wondered if bodies could act upon light at a distance, bending its rays [4]. Still in the 18th century, Henry Cavendish, based in Newtonian mechanics and corpuscular theory of light, would have made a computation of light deflection, although never published at the time [5]. Some years later, in the early 19th century, Johann Georg von Soldner computed and published the value of 0. 84 arcseconds for the bending of a light ray passing near the surface of the Sun. von Soldner wrote that if one could observe stars close enough to the Sun, one would have to take this into account, but he concluded that this could not be done, missing the thought that this could be possibly verified during a total solar eclipse [6].

Einstein, most probably not aware of any of these previous computations, working in Prague in 1911, obtained the numerical value of 0.83 arcseconds for the deflection of a light ray grazing the surface of the Sun. Einstein communicated with the astronomer Erwin Finlay-Freundlich, of the Berlin Observatory, about the possibility of verifying experimentally this light-bending effect. Giving full consideration to this matter, Freundlich even thought that such an effect could be measured during daylight, an idea that has been abandoned, until the birth of radio astronomy.

Freundlich also believed that light bending could be measured using the existing photographic plates taken during previous total solar eclipses. With that aim, in September 1911, Freundlich wrote to Charles Dillon Perrine, who had obtained many of such plates, while working at Lick Observatory, in the USA. Since 1909, Perrine was the Director of the Córdoba Observatory, which, at that time, was the Argentinean National Observatory.

---

[1] Faculdade de Física, Universidade Federal do Pará, 66075-110, Belém, PA, Brazil.
[*] crispino@ufpa.br
[2] Observatorio Astronómico de Córdoba - Museo Astronómico, Laprida 854, X5000BGQ Córdoba, Argentina.

When Freundlich's letter arrived at the Córdoba Observatory, Perrine was absent, being in a trip to Europe to participate of a congress of astronomers in Paris. Freundlich ended up by meeting Perrine in the Berlin train station, in October 1911. In that occasion, Freundlich asked to Perrine about the possibility of verifying light bending in the photographic plates taken by him, in the search for intramercurial planets, through the measurement of the displacement of the stars near the eclipsed Sun. Perrine answered to Freundlich that, because of the small fields and short exposures, as well as due to the fact that the Sun was not in a the center of those photographic plates, which were kept in the Lick Observatory, that should not be possible [7].

After this conversation with Perrine, Freundlich wrote to William Wallace Campbell, Director of the Lick Observatory, asking copies of the eclipse plates obtained by Perrine. Campbell replied to Freundlich in early 1912, informing that Lick Observatory actually had eclipse plates taken by Perrine during the total solar eclipses observed in Sumatra (1901), Spain (1905) and Flint Island (1908), explicitly mentioning several hundred of star images recorded in the Flint Island plates. In June of that same year Campbell sent a selection of copies of those eclipse plates to Freundlich, mainly those in which the Sun was in or near one edge. However, Freundlich could not obtain any conclusion about light bending from those Lick images, as Perrine had anticipated to him when they met in Berlin.

Back in Córdoba, in 1912, Perrine received a letter from Freundlich asking if he could, during the total solar eclipse of October 10 of that year, obtain photographic plates to verify the possible displacement of the star images due to the Sun's gravitational field. Campbell also wrote a letter to Perrine wondering if he would not take the opportunity of the 1912 solar eclipse to go to Brazil to measure light bending, being so kind as to offer to lend the Lick Observatory eclipse apparatus used to photograph stars near the Sun. Indeed, for the total solar eclipse of October 1912, Perrine organized an expedition of the Córdoba Observatory to Brazil, bringing with him, together with other instruments, the intramercurial camera lenses loaned by the Lick Observatory.

Apart from this Argentinean expedition, other international commissions went to Brazil, namely, British, French and Chilean, joining the expedition organized by the Brazilian National Observatory, leaded by its director Henrique Charles Morize (Fig. 1). A second Argentinean expedition, from the La Plata Observatory, also went to Brazil. Among all these, the only expedition that aimed to measure light bending was Perrine's. The others were mainly concerned with properties of the solar corona.

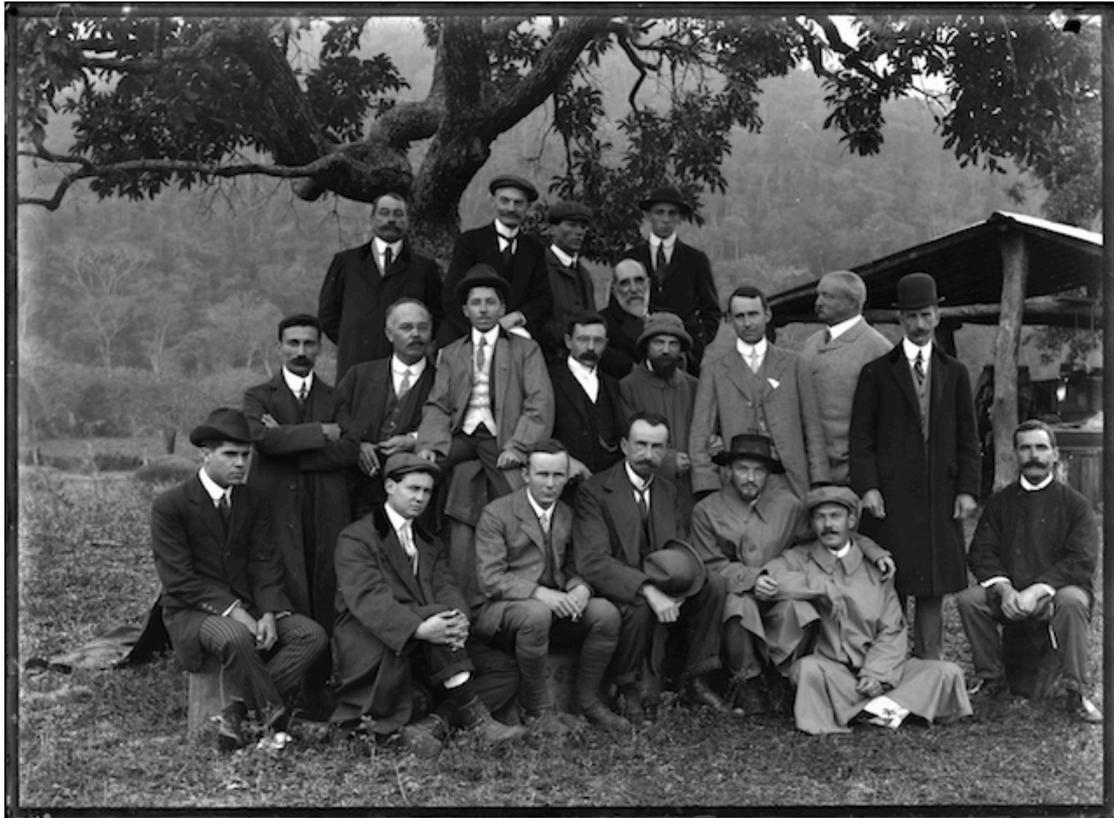

Fig. 1 | Astronomers gathered in Passa Quatro, in the State of Minas Gerais, Brazil, for the observation of the October 1912 total solar eclipse. The director of the Brazilian National Observatory, Henrique Charles Morize is the fourth (from left to right) seating in the front (first) row. The British Charles Davidson, Arthur Eddington and John Atkinson are the fourth, sixth and seventh, respectively, standing in the middle (second) row. Courtesy of Biblioteca do Observatório Nacional, Rio de Janeiro, Brazil.

The Brazilian, British and French expeditions stayed in a farm in Passa Quatro, in the State of Minas Gerais. Arthur Stanley Eddington and Charles Rundle Davidson composed the British commission, assisted by John Jepson Atkinson (Fig. 1). Perrine set camp for the Córdoba expedition in Cristina (Figs. 2 and 3), another location in Minas Gerais State. The site chosen by the Chilean expedition was the surroundings of Cristina as well, and the La Plata expedition stayed in Alfenas, also located in Minas Gerais.

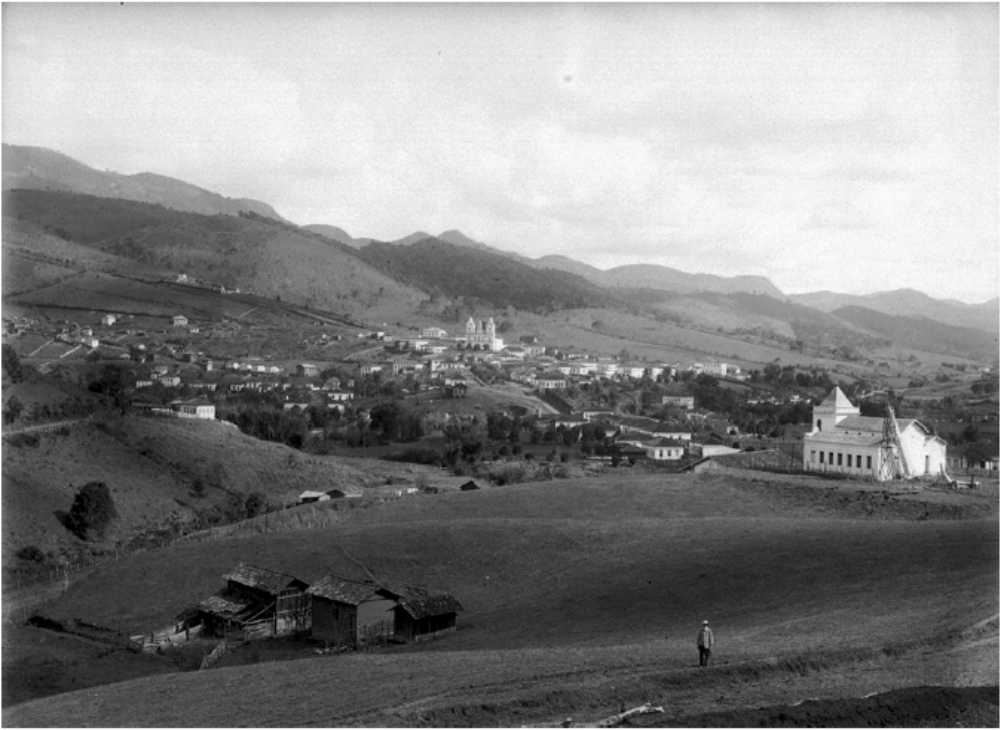

Fig. 2 | General view of Cristina, in Minas Gerais, Brazil, taken in October 1912. The Córdoba Observatory settlement appears in the right. Courtesy of Observatorio Astronómico de Córdoba - Museo Astronómico, Córdoba, Argentina.

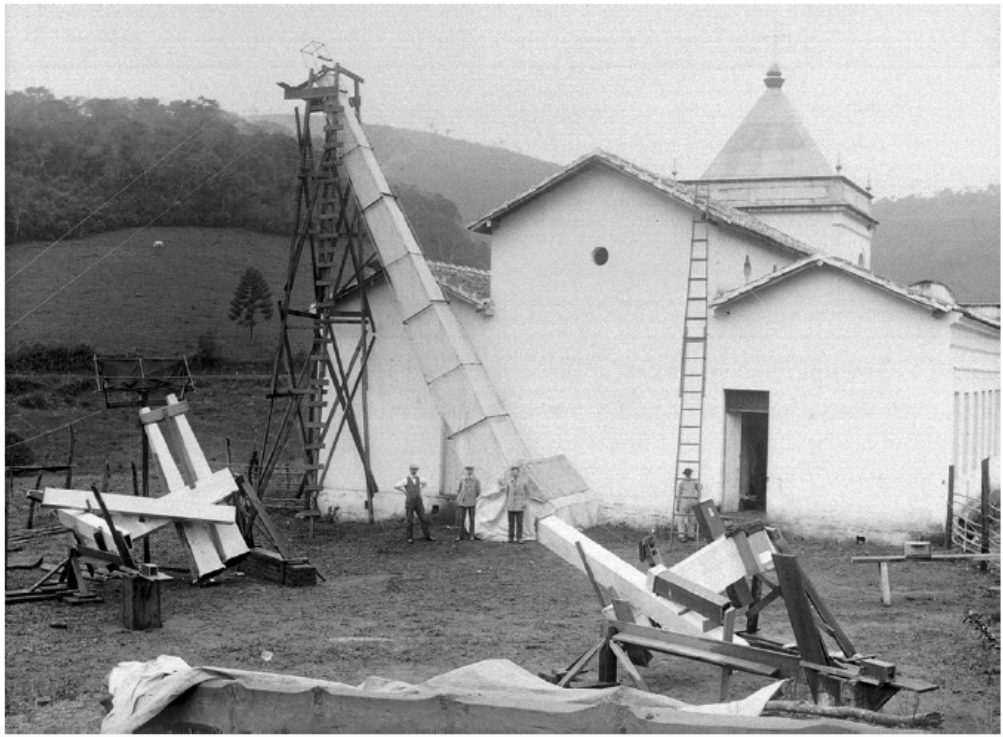

Fig. 3 | Córdoba Observatory settlement in Cristina, Brazil, for the observation of the October 10, 1912 eclipse. Courtesy of Observatorio Astronómico de Córdoba - Museo Astronómico, Córdoba, Argentina.

Perrine and Eddington first met each other in England. They met again at least twice in Rio de Janeiro, before going to their different observation sites in Minas Gerais. Eddington had dinner in Rio de Janeiro with Perrine, together with his three assistants from the Córdoba Observatory, namely the astronomer Enrique Chaudet, the mechanical James Oliver Mulvey and the photographer Roberto Winter. Whether they talked about Einstein prediction of light deflection by the Sun's gravitational field is an open question. If this happened, it may have been the first time Eddington heard about this matter.

Using Perrine's words to describe what happened in Brazil, the astronomers "suffered a total eclipse instead of observing one". It happened that heavy clouds and rain impeded any photograph of the eclipsed Sun to be taken in Brazil in1912 (Fig. 4). Still according to Perrine description, the intermittent rain lasted in his observation site for more than four days, including the eclipse in between. Therefore, light bending verification had to be postponed for further opportunities, the next one being the total solar eclipse of August 21, 1914, visible in the extreme north of America, as well as in some parts of Europe and Asia.

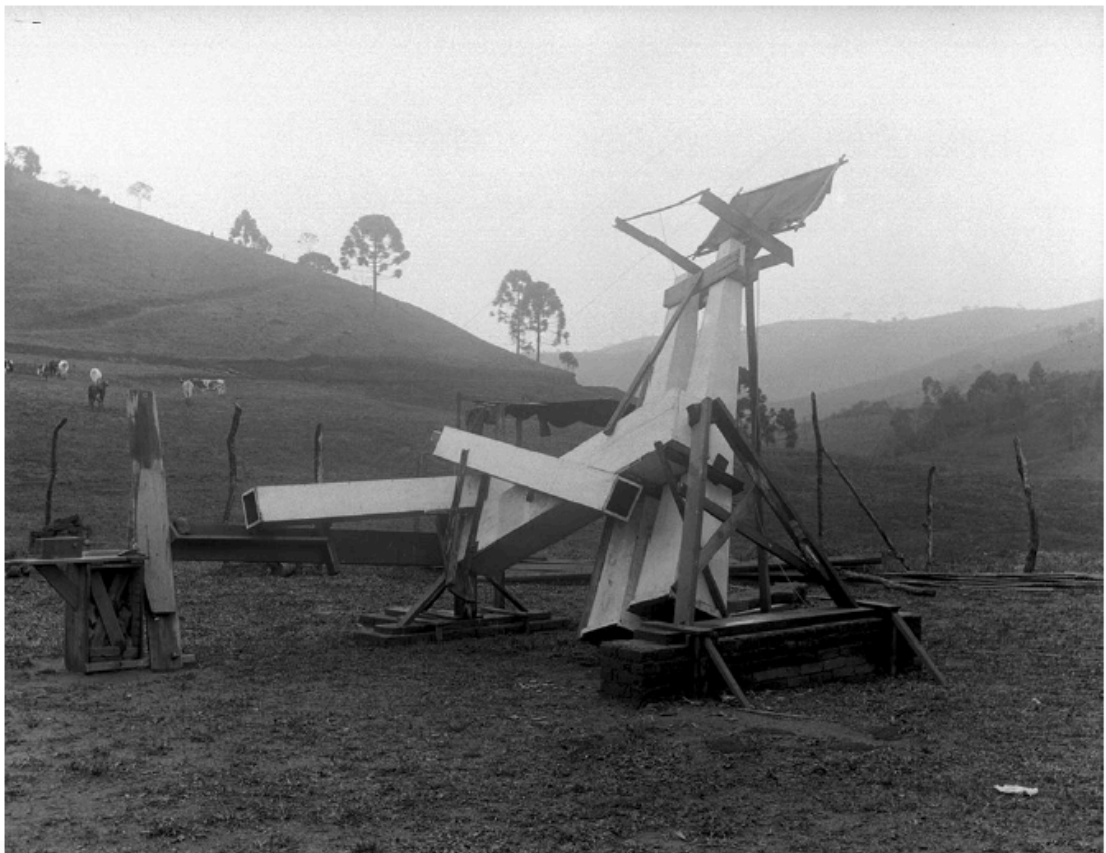

Fig. 4 | Intramercurial cameras brought by Perrine to measure light bending by the Sun in Cristina, Brazil, in 1912. Courtesy of Observatorio Astronómico de Córdoba - Museo Astronómico, Córdoba, Argentina.

Among the expeditions organized to observe this 1914 eclipse, stand out the Córdoba (Fig. 5) and the Berlin Observatory ones sent to Crimea, and the Lick Observatory one sent to Ukraine (Fig. 6). Essential parts of the intramercurial telescopes were sent by Perrine to Freundlich, and the latter was supposed to complete the cameras and join Perrine's expedition in Crimea. However, World War I started, related to the assassination of the Austro-Hungarian Archduke Franz Ferdinand, in Sarajevo, on July 28, 1914, just in time to impose serious difficulties to the traveling astronomers, most significantly to the German Freundlich. Therefore, in an expedition that Einstein himself helped to raise funds, Freundlich could not perform any measurement.

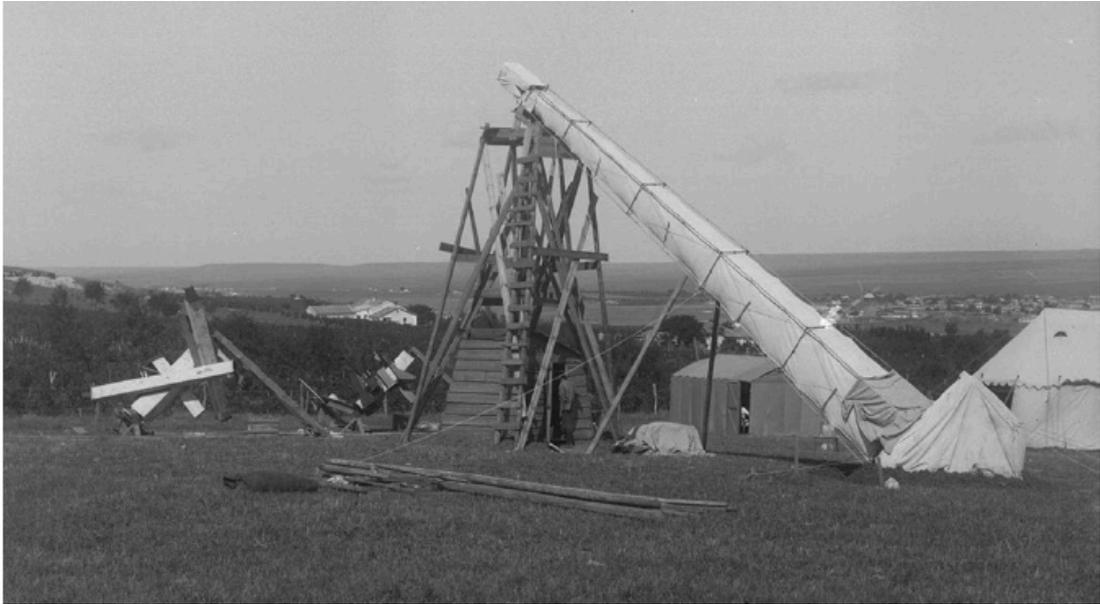

Fig. 5 | General view of the Córdoba Observatory settlement, near the city of Feodosia, which appears in the background. Courtesy of Observatorio Astronómico de Córdoba - Museo Astronómico, Córdoba, Argentina.

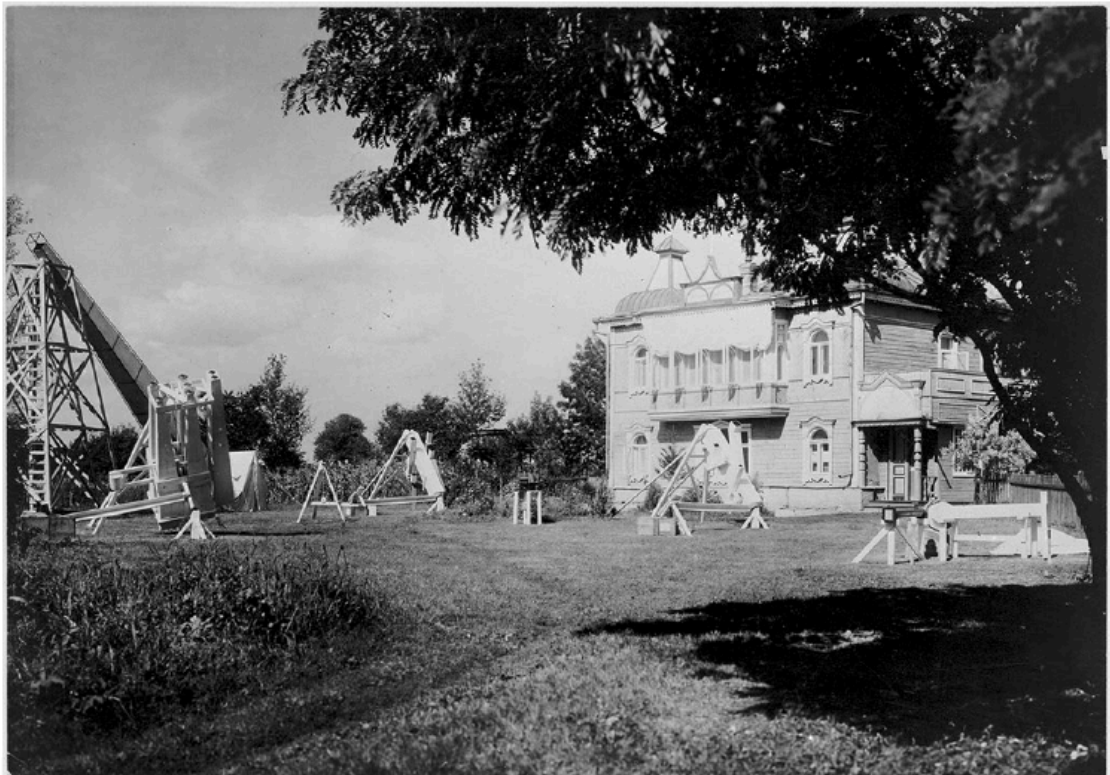

Fig. 6 | Lick Observatory settlement at Brovary, Imperial Russia (presently Ukraine), in 1914. Courtesy of Mary Lea Shane Archives, University of California, Santa Cruz, USA.

Through clouds, the Córdoba team was able toobtain some photographs (Fig. 7), but no measurement of star positions could be performed. Campbell and Heber Doust Curtis, from Lick Observatory, in Brovary, Ukraine, having their own intramercurial cameras (Fig. 8), were clouded out, and obliged to return to America leaving behind their instruments.

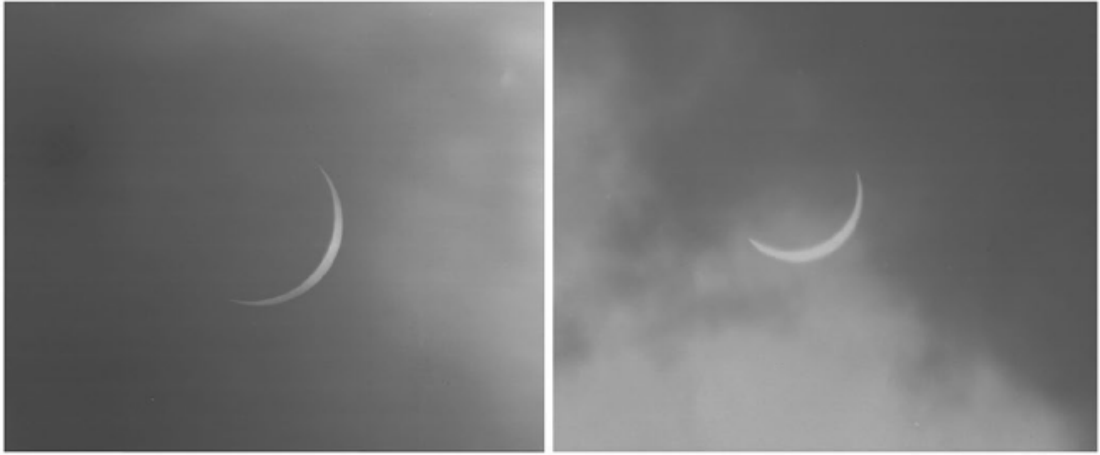

Fig. 7 | Photographs taken by the Córdoba team during the solar eclipse of 1914, in Crimea. Courtesy of Observatorio Astronómico de Córdoba - Museo Astronómico, Córdoba, Argentina.

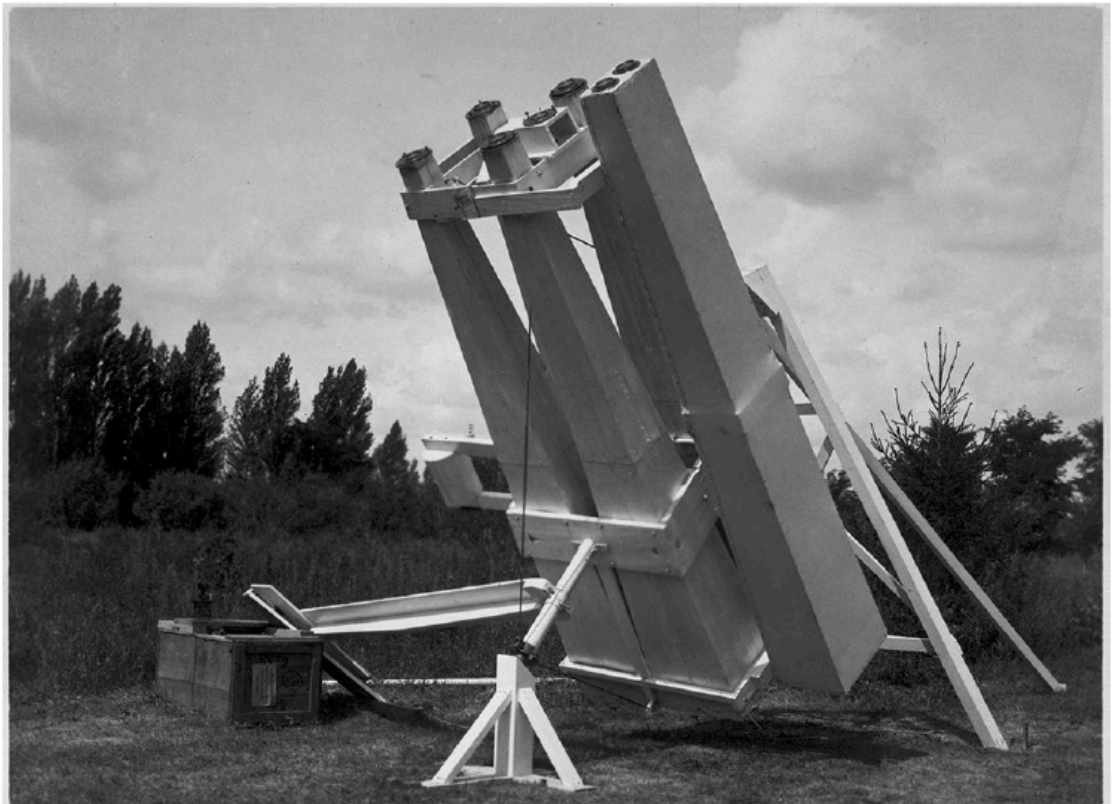

Fig. 8 | "Einstein-Vulcan" or intramercurial cameras mounted by the Lick Observatory team at Brovary, in 1914. Courtesy of Mary Lea Shane Archives, University of California, Santa Cruz, USA.

The following year happened to be crucial for gravitational light bending investigation, since, in November 1915, Einstein conceived the final form of the General Theory of Relativity, obtaining the correct result for light deflection by the Sun, and equal to twice the one he obtained back in 1911.

Although some solar eclipses happened in the meantime, the first experimental confirmation of General Relativity had to wait until the end World War I, for the 1919 Greenwich expedition observations conducted in Sobral (Brazil), by Andrew Claude de la Cherois Crommelin together with Davidson [8], and in Principe Island (African Western Coast), by Eddington together with Edwin Turner Cottingham [9]. Perrine, who has been the first astronomer that tried to measure light bending during a total solar eclipse, could not get the appropriate funding to realize his expedition, as planned, for the 1919 eclipse.

Curiously enough, if one of those 1912 or 1914 frustrated eclipse attempts had been successful, instead of proving that Einstein's result (current at the time) was right, the first measurement of light deflection by the Sun's gravitational field would have had the opposite outcome, and this part of the History of Science would have been surprisingly different.

**Acknowledgements**

L. C. acknowledges partial support from Conselho Nacional de Desenvolvimento Científico e Tecnológico (CNPq) and Coordenação de Aperfeiçoamento de Pessoal de Nível Superior (CAPES)—Finance Code 001, from Brazil, as well as the European Union's Horizon 2020 research and innovation programme under the H2020-MSCA-RISE-2017 Grant No. FunFiCO-777740.